\documentclass[runningheads]{llncs}
\usepackage[utf8]{inputenc}
\usepackage[T1]{fontenc}
\usepackage{graphicx,booktabs,amsmath,url,float} 
\usepackage{hyperref}
\usepackage{tabu}                      
\usepackage{booktabs}
\usepackage{lipsum}
\usepackage{mwe}                       
\usepackage{stfloats}  
\usepackage{csvsimple} 
\usepackage{array, ragged2e, amsmath}
\usepackage[backgroundcolor=yellow]{todonotes}

\usepackage{amsmath}
\usepackage[misc]{ifsym}
\usepackage[utf8]{inputenc}

\usepackage{mwe}
\usepackage{comment}
\usepackage{booktabs}  
\usepackage{multirow}  
\usepackage{graphicx}
\usepackage{pdflscape}
\usepackage{subcaption}
\usepackage{mathptmx}                  

\begin{document}

\title{Automated Visualization Makeovers with LLMs}

\author{%
  Siddharth Gangwar\inst{1} \and
  David A. Selby\inst{2} \and
  Sebastian J. Vollmer\inst{1,2}%
}

\institute{%
  University of Kaiserslautern–Landau (RPTU)\\
    \email{%
    \{byr34xub\}@rptu.de%
  }
  \and
  Department of Data Science and its Applications, German Research Center for Artificial Intelligence (DFKI)\\
  \email{%
    \{david\_antony.selby, sebastian.vollmer\}@dfki.de%
  }%
}
\date{}  

\titlerunning{Automated Visualization Makeovers with LLMs}
\maketitle

\begin{abstract}
    Making a `good' graphic that accurately and efficiently conveys the desired message to the audience is both an art and a science, typically not taught in the data science curriculum. `Visualisation makeovers' are exercises where the community exchange feedback to improve charts and data visualizations. Can multi-modal large language models (LLMs) emulate this task? Given a plot in the form of an image file, or the code used to generate it, an LLM, primed with a list of visualization best practices, is employed to semi-automatically generate constructive criticism to produce a `better' plot.
    Our system is centred around prompt engineering of a pre-trained model, relying on a combination of user-specified guidelines and any latent knowledge of data visualization practices that might lie within an LLM's training corpus.
    Unlike other works, the focus is not on generating valid visualization scripts from raw data or prompts, but on educating the user how to improve their existing data visualizations according to an interpretation of best practices.
    A quantitative evaluation is performed to measure the sensitivity of the LLM agent to various plotting issues across different chart types.
    We make the tool available as a simple self-hosted applet with an accessible Web interface.

  A copy of this paper and dataset is available at \url{https://OSF.io/4tb87}
\end{abstract}


\graphicspath{{figs/}{figures/}{pictures/}{images/}{./}} 

\section{Introduction}

Clarity and precision in data visualization is essential for effective communication across fields, but many practitioners lack formal training in design best practices, which are constantly evolving and not always easily taught in a traditional classroom setting \cite{bach_education_2024}.
Community-driven initiatives such as \href{https://makeovermonday.co.uk/}{\#MakeoverMonday} \cite{kriebel_makeovermonday_2018} exemplify efforts to enhance visualizations by reimagining existing charts to improve their effectiveness and adherence to stylistic guidelines.
Building on this, we explore the capabilities of multimodal large language models (LLMs) to assess and refine visualizations autonomously.
We propose a system that accepts charts as inputs (either as images or code representations) and identifies `grammatical' errors or design rule violations, such as inappropriate use of dual axes; or `style' errors, such as misuse of 3-d effects. The system identifies these issues and provides target suggestions for improvement.

While prior research has explored the capabilities of LLMs in generating exploratory visualizations from data, detecting errors within textual inputs or writing data analysis code, our approach focusses on the critical evaluation and improvement of existing visualizations.

Recent advances in multimodal LLMs, such as GPT-4 and Gemini, have made it possible to reason about both visual and textual inputs using a unified interface. These models~\cite{gpt4,gemini} can interpret visual structure, recognize design elements, and provide text-based explanations or code output, making them particularly promising for visualization critique tasks. However, their effectiveness in this domain depends heavily on prompt design, rule specificity, and how context is conveyed. Previous studies~\cite{lo_how_2025} have evaluated such models' ability to detect misleading or invalid chart elements, but have not emphasized correction, educational feedback, or modular extensibility.

In this work, we present a system that combines structured prompting with chart-type-specific design rules to identify common visualization issues and offer actionable feedback. Our system supports both image and code inputs, making it usable across a wide range of tools, from scripting libraries such as Matplotlib to GUI-based platforms like Tableau or Excel. When a chart image or chart script is submitted, the system first identifies the chart type, loads relevant visualization guidelines, and queries a multimodal LLM to detect violations and propose corrections.

To evaluate the system, we created a synthetic dataset of charts with known issues and manually curated labels. We report multi-label classification metrics, including precision, recall, and F1-score per error type, as well as aggregate statistics such as Mean Absolute Error (MAE) for predicted issue count. Our results show that the system performs reliably on objective structural errors, while stylistic issues remain more ambiguous and difficult for LLMs to detect consistently.

Our contributions are:
\begin{itemize}
    \item A multimodal system that analyzes both image- and code-based charts and identifies design issues using chart-specific rules. 
    \item A prompting framework that supports visual reasoning using LLMs and provides natural language explanations of detected issues.
    \item A quantitative evaluation on a labeled synthetic dataset, demonstrating the system’s strengths in detecting objective chart errors.
\end{itemize}

\section{Related Work}

\subsection{Rule-Based and Heuristic Systems}

A number of systems have been developed to assess the quality of data visualizations by enforcing predefined design rules. Early rule-based tools, such as VisuaLint~\cite{hopkins_visualint_2020} annotates charts with sketch-style overlays to highlight issues such as truncated axes or missing legends. VizLinter~\cite{vizlinter} formalizes best practices using Answer Set Programming (ASP) in Vega-Lite specifications and corrects violations with a linear programming module. These tools are typically tied to specific chart grammars and rely heavily on manually encoded heuristics. Our work builds on the idea of rule-driven analysis, but extends it to multimodal contexts—allowing critique from visual inputs and dynamically loading chart-type-specific rules during inference.

\subsection{Explanation and Interaction Approaches}

In parallel, research has explored how to effectively communicate visualization errors to users. Lo et al.~\cite{lo_why_2024} compared six explanation strategies, including annotation, highlighting, redrawing, and correction, through a large-scale crowdsourced study. Their findings emphasized that clear and interactive explanations, such as explorable visual feedback, enhance user understanding and trust. Shen et al.~\cite{shen_towards_2023} provide a comprehensive survey of natural language interfaces for data visualization, highlighting the growing interest in systems that allow users to query or refine visualizations using conversational input.

\subsection{LLMs for Visualization Generation and Captioning}

Recent work has applied large language models (LLMs) to tasks like visualization generation and captioning. LIDA~\cite{dibia_lida_2023} uses a multi-stage pipeline with summarization, goal setting, and visualization generation to produce infographics directly from data. Unlike our focus on critique and error detection, LIDA focuses on visualization generation from text summaries and does not support post-hoc analysis of existing visualizations. Other work~\cite{liew_captions_2022} has evaluated LLMs' performance in generating descriptive captions for charts. VisEval~\cite{chen_viseval_2024} introduced a benchmark for assessing LLM-generated visualizations across dimensions such as validity, legality, and readability. Bavisitter~\cite{choi2023bavisitter} incorporates design guidelines directly into the LLM workflow to provide natural language suggestions that improve visualizations generated by the model. These systems demonstrate the expressive power of LLMs, especially in natural language generation. However, they typically operate on structured text or code and do not focus on post-hoc critique from visual inputs, which is the core of our system.

\subsection{LLMs for Critique and Feedback}

Beyond generation, LLMs have been studied for their ability to provide post-hoc feedback on visualization quality. Kim et al.~\cite{kim_chatgpt_2024} evaluated ChatGPT’s design feedback on real-world visualization questions from the VisGuides forum, finding that while ChatGPT responses matched humans in clarity and coverage, they sometimes lacked contextual depth. Lo and Qu~\cite{lo_how_2025} investigated whether multimodal LLMs could detect misleading charts from bitmap images, using general prompts to identify 21 common visualization issues. Their results revealed strong potential for visual reasoning, but also highlighted issues with prompt sensitivity and false positives. Their evaluation leveraged a curated dataset introduced in earlier work~\cite{lo_misinformed_2022}.
Our work draws inspiration from this line of research but expands the scope from misinformation detection to a broader spectrum of visualization design issues. In contrast to general prompts, we use modular chart-type-specific prompts and rules to improve detection consistency and offer actionable feedback.

Together, these prior systems establish that both rule-based and LLM-based methods can contribute to visualization critique. However, most either assume access to structured code or lack modular reasoning over chart types. Our system aims to unify these strengths—combining rule-driven critique with the visual reasoning capabilities of multimodal LLMs—and to support both image and code inputs with extensible evaluation pipelines.

\section{Goals \& Requirements}

Unlike other code generation tasks, it is not enough to evaluate whether or not a visualization script successfully compiles.
Indeed the interface need not (and perhaps \emph{should} not) be code based, as the system ideally is not only language-agnostic but also able to handle graphics whose source code is unavailable or which have been created using GUIs such as \textit{Tableau} or \textit{Microsoft Excel}.
Moreover, as an educational tool to teach good practice, it may be better to provide constructive suggestions for improvement and offer general advice, rather than simply generating an updated graphic or script.

Nevertheless, to evaluate the efficacy of such a tool reproducibly, a quantitative evaluation criterion is desired.
However, while examples of real-world visualizations (and the code that generated them) may be available in various repositories, such as Kaggle or Papers with Code, they do not necessarily represent best practice; a structured open database of `good' and `bad' visualization practices, accompanied with real-world training examples, is not readily available.
Two possible solutions are available: (a) the labour-intensive task of trawling the Web for representative examples of different visualizations, including common errors or combinations thereof, and manually annotating them; or (b) for a given set of chart types and possible charting problems, generating a dataset of example plots with these combinations of characteristics in various formats.
The second possibility can be accelerated via LLMs, with a human-in-the-loop to verify the generated plots exhibit the desired features (ideally without inadvertently introducing unlabelled issues), by generating scripts to visualize toy datasets.

Given this context, the goal of our system is to enable automated chart critique from both rendered images and code-based inputs. It should employ chart-type detection to dynamically apply tailored rule sets, ensuring that best practices are evaluated in the context of the specific chart format, whether bar, line, pie, or others. Rather than providing binary correctness judgments, the system identifies specific design issues and generates natural language explanations, and if code is provided, corrected code that adhere more closely to good visualization practices.

To support these goals, the system must handle multimodal inputs and detect multiple concurrent error types per visualization. It should support chart-specific reasoning based on structured rule sets, use LLMs not only to perceive visual elements but also to evaluate their appropriateness, and produce outputs that are understandable to users. Finally, performance must be evaluated using multi-label metrics such as precision, recall, and F1-score, using a dataset that reflects a variety of error conditions across multiple chart types.

\begin{figure}[H]
  \centering
  \includegraphics[width=\linewidth]{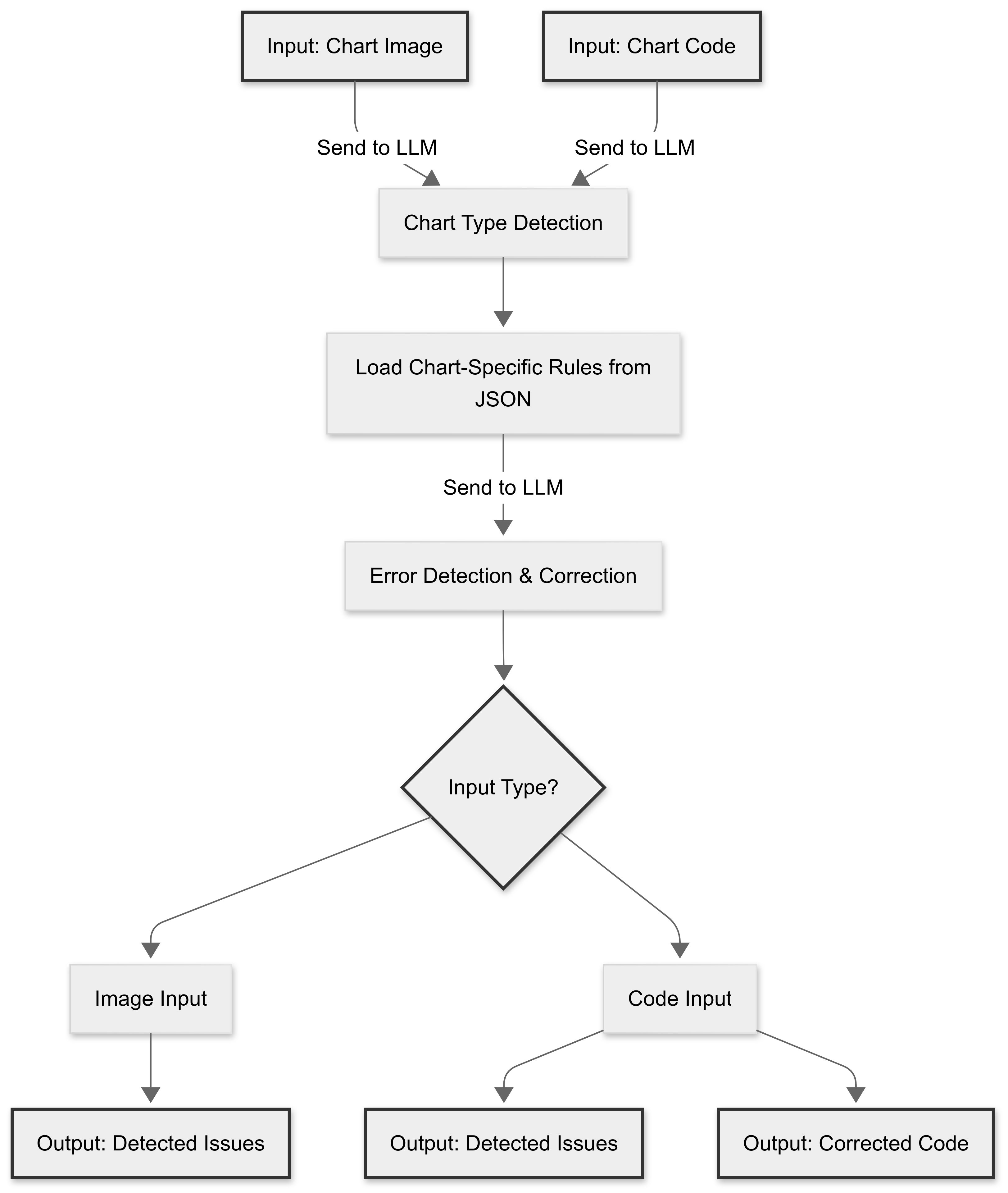}
  \caption{Workflow of the system: Inputs (as images or code) are processed through chart-type detection, rule-based prompting, and error analysis via an LLM. Outputs include detected issues and (optionally) corrected code.}
  \label{fig:workflow}
\end{figure}

\section{Implementation}

We demonstrate a working prototype of our system designed to analyze visualizations for common design issues and suggest improvements. The interface supports two modes of input: users may either upload an image of a chart (e.g., a screenshot or exported plot) or paste the corresponding visualization code.

\textbf{Prompt Structure.}
The system uses a modular, multi-stage prompting pipeline tailored to the chart type and input mode (image or code). Below is a breakdown of each stage:

\begin{description}
  \item[1. Chart Type Detection:] 
  The model is prompted to identify the chart type from the input. LLM Prompt: \textit{Detect the type of chart represented... Respond with only the chart type name (e.g., Bar Plot, Line Plot, etc.).}

  \item[2. Threshold Evaluation:] 
  Based on the identified chart type, the model is asked to extract relevant properties (e.g., number of categories, , number of lines, axis range) and assess them against predefined thresholds.

  \item[3. Rule Loading and Application:] 
  For each chart type, visualization rules are stored in a structured JSON file. These include best practices and constraints (e.g., “No more than 7 pie slices”, “Avoid dual axes for line charts”). The system loads and applies these rules to the current chart context.

  \item[4. Issue Detection \& Feedback:] 
  The LLM analyzes the chart in context of the loaded rules and thresholds, identifying design flaws. It generates natural-language feedback describing each issue and how it violates the rule.

  \item[5. Code Correction (optional):] 
  If the input is code (e.g., Python/Matplotlib), the LLM may also generate a corrected version that fixes the issues.

  \item[6. Output Display:] 
  The final feedback, including detected issues and suggestions, is presented in the web interface. For image inputs, annotations accompany the original chart; for code inputs, side-by-side code suggestions are shown.
\end{description}

The frontend interface is web-based and features a clean, two-column layout: the left pane displays the user’s input (image or code), and the right pane shows the system's analysis and output. If an image was uploaded, it is shown in place of the code editor for context. The interface includes buttons to submit or clear inputs, and supports real-time feedback display.

The backend is implemented in Python using Flask, with prompt-based interaction handled through the OpenAI GPT-4o API. Image inputs are encoded and sent along with tailored prompts, while code inputs are passed as-is. The rule sets for each chart type are defined in an external JSON file and loaded dynamically based on LLM-detected chart types. The frontend is developed using HTML, CSS, and JavaScript.

\begin{figure}[H]
  \centering
  \includegraphics[width=\linewidth]{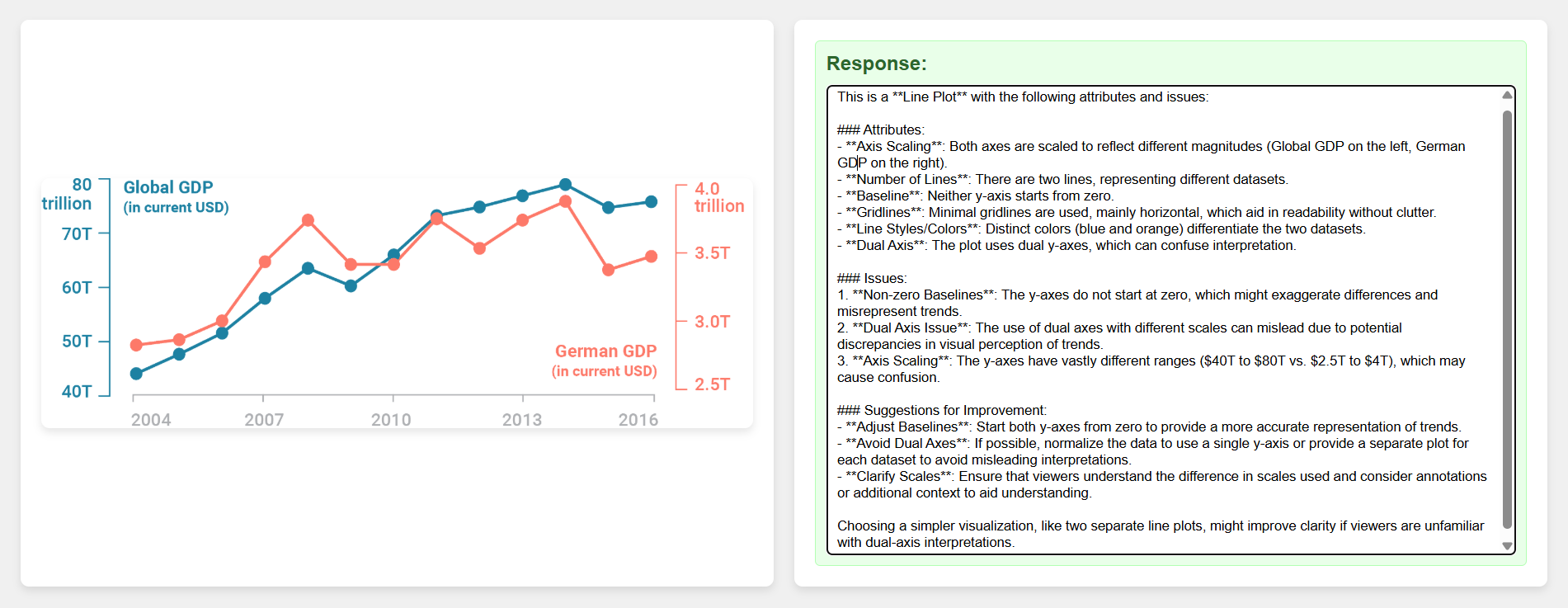}
  \caption{%
  	Demonstration: Uploaded chart (sourced from \cite{gdp}) is analyzed by the system. On the right, the interface shows detected issues such as non-zero baseline and dual-axis issue, with suggested fixes.%
  }
  \label{fig:teaser}
\end{figure}

Figure~\ref{fig:teaser} illustrates an example demonstration. The user uploads a chart image (in this case, a line plot with a non-zero $y$-axis baseline, dual-axes and improper scale). The system successfully identifies both issues, provides natural language feedback, and lists explanations aligned with best practice guidelines.

\section{Evaluation}
To assess the accuracy and robustness of our system, we conducted a quantitative evaluation based on a controlled dataset of chart images exhibiting known visualization issues. Our evaluation centers on the detection of visual issues. Although the system provides feedback and suggested corrections, their effectiveness was not evaluated in a user-facing study. The evaluation was designed to answer three research questions: 
\begin{itemize}
    \item(RQ1) How well does the system detect specific error types and which types are most challenging? 
    \item(RQ2) How far off, on average, is the predicted number of errors from the ground truth, and does the system tend to overestimate or underestimate? 
    \item(RQ3) How does the system’s performance vary on images containing a single error versus multiple errors?
\end{itemize}

\vspace{1mm}
\textbf{Dataset.}  
We created a synthetic dataset of 72 visualization images encompassing 12 distinct error types: \textit{Improper Scale or Axis Range, Non-Zero Baselines, Overuse of Gridlines, Dual Axis Issues, Inconsistent Bar Widths, Overuse of 3D Effects, Inappropriate Colour Choices, Too Many Slices in Pie Charts, Missing or Inadequate Labels/Legends, Overlapping Data Elements, Uneven or Inconsistent Tick/Time Intervals, Poor Category Ordering}. For each error type, we generated 6 example plots using varied chart types such as bar, line, and pie. Roughly 42\% of the dataset (30 images) was designed to contain more than one error, while the remaining 42 images each had a single issue. Each image was manually annotated to indicate both the presence of specific error types and the total number of issues, forming the ground truth labels.
Examples are given in \autoref{tab:error_types}.

The evaluation was conducted on image inputs only, as the goal was to assess the system’s visual error detection capabilities. Each image was processed once, and the predicted errors were compared to the ground truth annotations.

\vspace{1mm}
\textbf{Metrics.}  
We evaluated the system using standard multi-label classification metrics: \textbf{precision}, \textbf{recall}, and $F_1$-\textbf{score}, computed per error type. Precision measures the proportion of correctly predicted instances out of all predictions made for a given error; recall measures the proportion of actual error instances that were successfully detected. The $F_1$-score is the harmonic mean of precision and recall.

To assess how well the system predicted the total number of errors per image, we also computed the \textbf{Mean Absolute Error (MAE)}.

\begin{table}[H]
    \centering
    \caption{Some of the major issues commonly encountered in visualization practices. Illustrations sourced from Tableau documentation \cite{tableau}.}
    \label{tab:error_types}
    \begin{tabular}{m{2.5cm}>{\raggedright\arraybackslash}m{3.5cm} >{\raggedright\arraybackslash}m{8.5cm} m{9.5cm}}
        \toprule
        Error Type & 
        Description & 
        Illustration \\
        \midrule
        Improper Scale or Axis Range & The scale or range of an axis is manipulated to exaggerate or minimize visual differences, potentially misleading the viewer about the significance of the data. & \includegraphics[width=8cm]{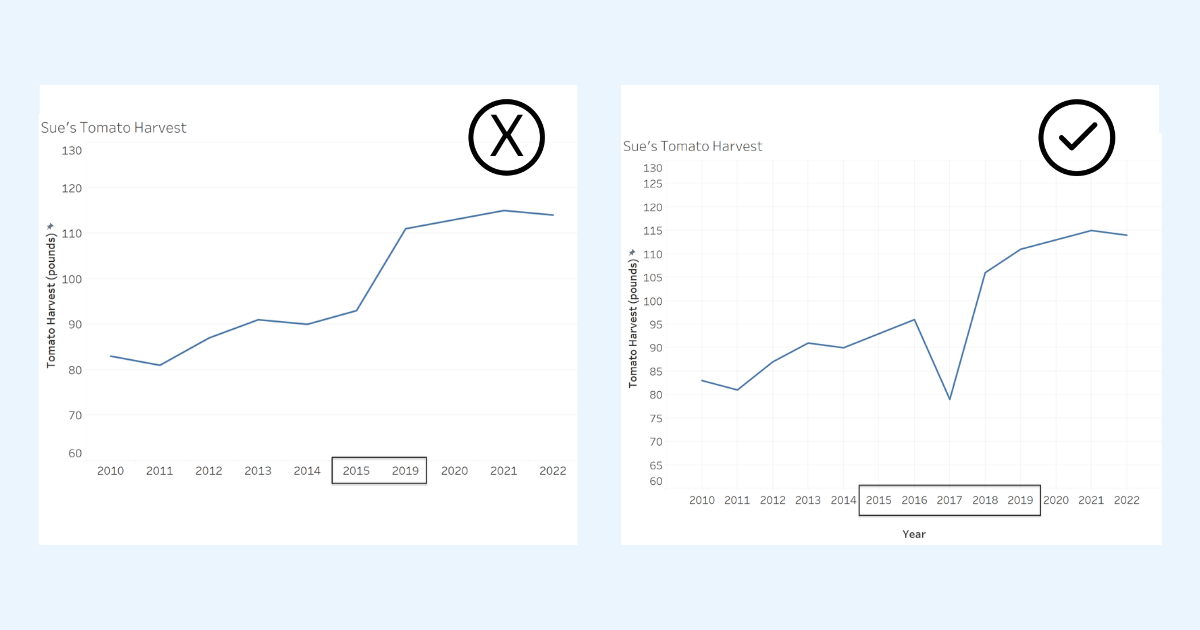} \\
        \midrule
        Non-Zero Baselines & An axis that does not start at zero can visually distort the magnitude of changes or differences between data points, making small differences appear larger. & \includegraphics[width=8cm]{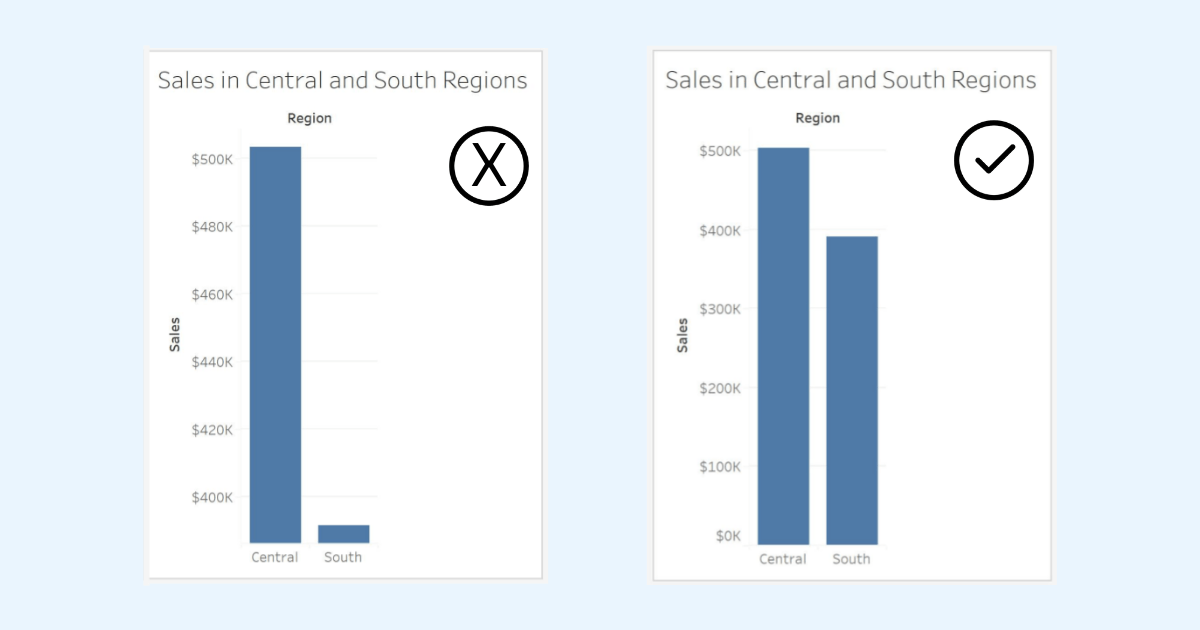} \\
        \midrule
        Dual Axis Issues & The use of dual axes with arbitrary scales can mislead readers by exaggerating or downplaying the relationship between two data series. & \includegraphics[width=8cm]{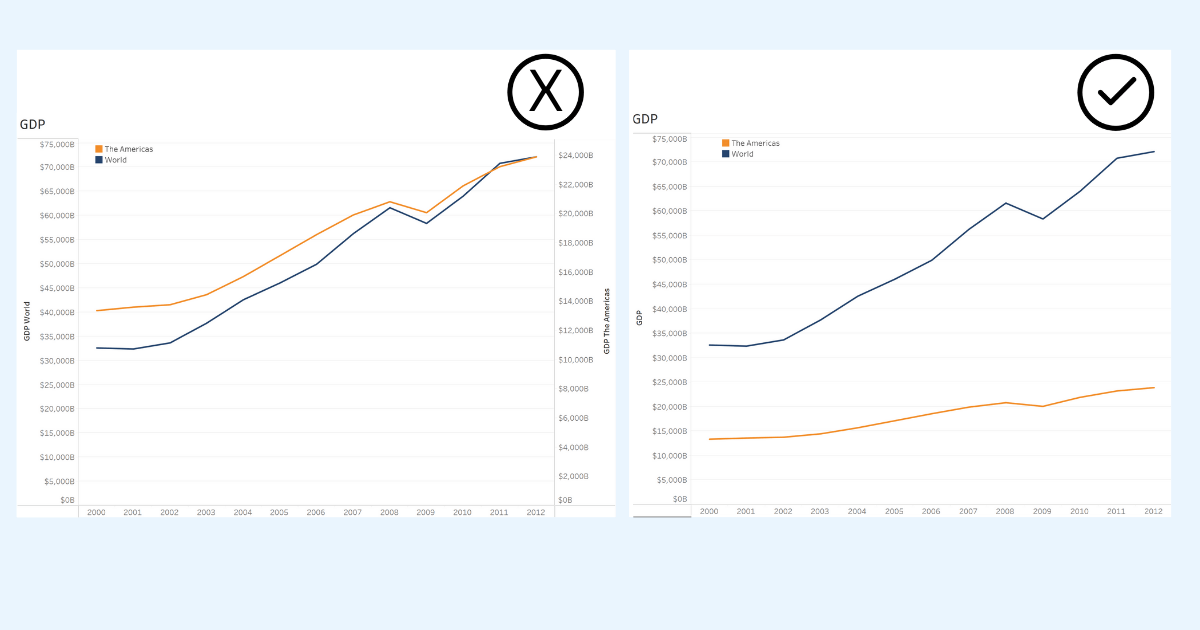} \\
        \bottomrule
    \end{tabular}
\end{table}

\section{Results}  
The system performed especially well in detecting error types with clearly visible and well-defined visual patterns. As shown in Table~\ref{tab:scores}, it achieved perfect F1-scores (1.00) for \textit{Non-Zero Baselines} and \textit{Dual Axis Issues}, and high scores for \textit{Too Many Slices}, \textit{Improper Axis Scaling}, and \textit{Inconsistent Bar Widths}. On the other hand, more stylistic or ambiguous error types such as \textit{Inappropriate Colour Choices} (F1 = 0.46) and \textit{Overlapping Data Elements} (F1 = 0.63) were more frequently misclassified.

\begin{table}[H]
  \centering
  \caption{Evaluation scores for each error type based on a single-pass run over each image in the dataset.}
  \label{tab:scores}
  \begin{filecontents*}{figs/scores.csv}
Error Type,Precision,Recall,F1-Score
Non-Zero Baselines,1.00,1.00,1.00
Dual Axis Issues,1.00,1.00,1.00
Too Many Slices in Pie Charts,1.00,0.83,0.91
Improper Scale or Axis Range,1.00,0.80,0.89
Inconsistent Bar Widths,1.00,0.67,0.80
Overuse of 3D Effects,1.00,0.67,0.80
Overuse of Gridlines,0.79,0.73,0.76
Missing/Inadequate Labels/Legends,0.70,0.78,0.74
Inconsistent Tick/Time Intervals,1.00,0.59,0.74
Poor Category Ordering,0.63,0.71,0.67
Overlapping Data Elements,0.50,0.83,0.63
Inappropriate Colour Choices,0.43,0.50,0.46
\end{filecontents*}
\csvreader[
    tabular= {> {\raggedright\arraybackslash}l >{\centering\arraybackslash}r >{\centering\arraybackslash}r >{\centering\arraybackslash}r}, 
    table head=\toprule Error Type & Precision & Recall & $F_1$ \\\midrule,
    late after line=\\,
    table foot=\bottomrule
  ]
  {figs/scores.csv}{}%
  { \csvcoli & \csvcolii & \csvcoliii & \csvcoliv }
\end{table}

The overall Mean Absolute Error (MAE) for the predicted error counts was \textbf{0.44}. This means that, on average, system’s prediction deviates from the ground truth by about 0.44 errors. Moreover, when we calculated the average difference between the predicted and actual error counts, we found that the system underestimates the number of errors by approximately \textbf{0.11} on average. This suggests that while the system’s count predictions are fairly accurate, there is a slight tendency for underestimation.

To answer RQ3, we divided the dataset into single-error and multi-error subsets. For single-error images, the MAE was 0.37, whereas for multi-error images it increased to 0.51, highlighting the challenge posed by overlapping issues.

\vspace{1mm}
In conclusion, our evaluation shows that the system is highly effective in identifying objective structural flaws but less reliable for more interpretive or stylistic issues. While it generally provides accurate feedback on individual error types, its performance on complex visualizations with overlapping issues can be improved.

\section{Conclusion \& Future Work}

We presented a modular, multimodal system that leverages large language models (LLMs) to automatically detect and explain visualization issues across both code-based and image-based inputs. By combining chart-type detection with structured, chart-specific design rules, the system provides interpretable feedback and, when applicable, generates corrected code to improve clarity and adherence to best practices.

Through a quantitative evaluation on a synthetic dataset of 72 annotated visualizations, the system demonstrated strong performance on well-defined structural issues, such as non-zero baselines and dual axes. However, it showed reduced accuracy on more stylistic or subjective issues like color usage and element overlap. These results underscore the potential of LLMs in visualization critique, while also revealing challenges in handling visual ambiguity and interpretive nuance.

This work also highlights several limitations. The current system assumes charts are self-contained and does not incorporate information about the underlying dataset. Its accuracy is influenced by the quality of the prompt engineering and visual input resolution. Furthermore, the evaluation relies on a curated, semi-synthetic dataset, which limits generalizability to uncontrolled, real-world inputs.

While the system is designed to support educational feedback and correction, our current evaluation focuses on the accuracy of issue identification. In future work, we discuss potential for expanded user-centered evaluation. Future directions also include extending the system to incorporate data-aware reasoning, expanding the rule base to cover more complex chart types and design heuristics, and improving visual robustness through computer vision integration.

\section{Acknowledgments}The authors wish to thank Professor Dr.\ Heike Leitte at RPTU, who provded valuable feedback during the development of this work.

\bibliographystyle{abbrv-doi-hyperref-narrow}
\bibliography{references}

\appendix 
\end{document}